\documentclass[twocolumn,prb,aps,amsmath,showpacs]{revtex4}
\usepackage{graphicx}% Include figure files

\usepackage{dcolumn}% Align table columns on decimal point

\usepackage{bm}% bold math
\usepackage{subfigure}
\usepackage{ltxtable, tabularx, longtable}

\begin{document}
\title{A DFT study of structural, dynamical properties and quasiparticle band structure of solid nitromethane}

\author{S. Appalakondaiah and G. Vaitheeswaran$^*$}
\affiliation{Advanced Centre of Research in High Energy Materials (ACRHEM),\\
University of Hyderabad, Prof. C. R. Rao Road, Gachibowli, Andhra Pradesh, Hyderabad- 500 046, India.}

\author{S. Leb\`egue}
\affiliation{Laboratoire de Cristallographie, R\'esonance Magn\'etique et Mod\'elisations (CRM2, UMR CNRS 7036), Institut Jean Barriol, Universit\'e de Lorraine, BP 239,\\
 Boulevard des Aiguillettes, 54506 Vandoeuvre-l\`es-Nancy, France.}

\date{\today}
\begin{abstract}
We report a detailed theoretical study of the structural, vibrational, and optical properties of solid nitromethane using first principles density functional calculations.
The ground state properties were calculated using a plane wave pseudopotential code with either the local density approximation (LDA), the generalized gradient approximation (GGA),
 or with a correction to include van der Waals interactions. Our calculated equilibrium lattice parameters and volume using a dispersion correction
are found to be in reasonable agreement with the experimental results. Also, our calculations reproduce the experimental trends in the structural properties
at high pressure. It was found to be a discontinuity in the bond length, bond angles and also a weaking of hydrogen bond strength in the pressure range from 10 to 12 GPa, picturing the structural transition from phase I to Phase II.
Moreover, we predict the elastic constants of solid nitromethane and found that the corresponding bulk modulus is in good agreement with experiments.
 The calculated elastic constants are showing an order of C$_{11}$$>$ C$_{22}$ $>$ C$_{33}$, indicating that the material is more compressible along the c-axis.
We also calculated the zone center vibrational frequencies and discuss the internal and external modes of this material under pressure. From this, we found the softing of lattice modes around 8 to 12 GPa. 
 We have also attempt the quasiparticle band structure of solid nitromethane with the G$_0$W$_0$ approximation 
 and found that nitromethane is an indirect band gap insulator with a value of the band gap of about 7.8 eV with G$_0$W$_0$ approximation. Finally, the optical
properties of this material, namely the absorptive and dispersive part of the dielectric function, and the refractive index and absorption spectra are calculated and the contribution of different
transition peaks of the absorption spectra are analyzed. The static dielectric constant and refractive indices along the three inequivalent crystallographic directions
 indicate that this material has a considerable optical anisotropy.
%\pacs{71.27+a, 71.18+y}
\end{abstract}
\maketitle
\section{Introduction}

Nitromethane is one of the simplest organic compound which belong to the class of secondary explosives. It is insensitive, safe to handle, used as a storable monopropellant
 and as an additive fuel for combustion engines. Because of its simple crystal structure compared with other secondary explosives, it has been used as a prototype
 material to understand the energetic materials by both experimentalists\cite{Trevino,Seminario,Cromer,yarger, Citroni,Hill, Miller,Ouillon, Pinan, OOuillon}
  and theoreticians.\cite{Sorescu, Rice,Byrd, Reed, Liu, Agarwal, zerili} Nitromethane is
 liquid at room temperature but the molecules condense to form a solid at around 4.2 K. Trivero et al\cite{Trevino} first reported X-rays and neutron
 diffraction experiments, and found that solid nitromethane crystalizes in an orthorhombic structure in the space group P2$_1$2$_1$2$_1$. Also, several experimental studies
 have been reported on the structural\cite{Cromer,yarger, Citroni} and the dynamical\cite{Hill, Miller, Ouillon, OOuillon, Pinan} properties of nitromethane at various pressures and temperatures. A considerable
 amount of theoretical  work has been performed using classical and quantum molecular dynamics simulations under various conditions.\cite{Mark, Alper, Rice, Sorescu, Riad} By
 using classical interatomic potential molecular dynamics simulations, Sorescu et al.\cite{Sorescu} predicted the structural properties of nitromethane crystals. The thermal decomposition of nitromethane was investigated using Car-Parrinello molecular dynamics simulations, which reveals that it
 undergoes a thermal decomposition at 2200 K with the cleavage of the C-N bonds.\cite{chang} The anisotropic nature of solid nitromethane was studied using pseudopotential approach by applying hydrostatic  pressure up to 20 GPa.\cite{Liu} Recently, Sorescu et al.\cite{dftd} and Landerville et al.\cite{Landeville} reported the structural properties of nitromethane 
 using respectively the parametrization of Grimme\cite{Grimme} and of Neumann and Perrin.\cite{Neumann}

\par Accurate and reliable calculations of electronic properties of secondary explosive materials with standard Density Functional Theory(DFT) are a difficult and
 challenging issue due to their complex crystal structure mainly controlled by van der Waals interactions. Also, the electronic band gap plays a major role in determining the impact
 sensitivity and detonation or defragmentation properties of explosive materials. It is well known that
 standard DFT using exchange-correlational functionals like the local density approximation (LDA) or the generalized gradient approximation (GGA)  always underestimate the
 band gap by about 30 - 40$\%$,\cite{Mel, Jones} therefore one has to use for instance the GW approximation to know the exact band gap value accurately.\cite{seb1, seb2}
 In the present study, efforts are taken to understand the structural and dynamical properties of solid nitromethane at zero and high pressures including van der Waals interaction.
 We also predict the elastic constants of this material to explore the stability and  detonation characteristics. In addition to these calculations,  quasiparticle band structure using the GW approximation
 and thereby the optical properties of the solid nitromethane are studied. The rest of the paper is organized as follows: in section II, we briefly describe the computational techniques
 while the results and discussions are presented in section III. Finally, a brief conclusion is given in section IV.

\section{Computational details}
The ground state properties are obtained using DFT in the Kohn-Sham formalism\cite{kohn} as implemented in the CASTEP simulation package\cite{Payne, Segall}.
 The Kohn-Sham orbitals of the valence electrons are expressed in a plane wave basis set and
  ultrasoft pseudopotentials are used to describe the electron-nuclei interactions. The LDA\cite{PPerdew} as well as the GGA schemes of Perdew and Wang\cite{Wang} (PW91) and
  of Perdew-Burke-Ernzerhof\cite{Perdew} (PBE) for the treatment of the exchange and correlation are used.
  Structural optimizations were performed by minimizing the forces on each atom until they were lower
 then 0.03 eV\AA$^{-1}$. A plane wave kinetic energy cutoff of 620 eV was used and the Brillouin zone  was
 sampled on a regular Monkhorst-Pack \cite{Monkhorst} grid of 6x5x4 points with a minimum spacing of 0.025 \AA$^{-1}$. The pseudopotentials were considered according to
 Vanderbilt's ultrasoft \cite{Vanderbilt} scheme with a number of valence electrons of 1 for hydrogen, 4 for carbon, 5 for nitrogen, and 6 for oxygen. To take into account vdW
 interactions which are oftenly present in molecular crystals, we have used various DFT-D methods as implemented in CASTEP. In the case of DFT-D calculations, the
 total energies of the system can be expressed as E$_{DFT-D}$=E$_{DFT}$+E$_{dispersion}$, where E$_{DFT}$ is the self consistent Kohn-Sham energy and E$_{dispersion}$
 is written as $ f(R_{ij}) \times C_6^{ij} \times R_{ij}^{-6} $  for each pair $(ij)$  of atoms separated by a distance $R_{ij}$. $f(R)$ is a damping function which is
 necessary to avoid divergence for small values of $R$, and  $C_6^{ij}$ is the dispersion coefficient for the atom pair $(ij)$. In the present work, different DFT-D approaches
 to treat vdW interactions were employed, notably the Ortmann, Bechstedt and Schmidt\cite{OBS} (OBS) correction to PW91, as well as the Tkatchenko and Scheffler\cite{TS} (TS) and
 Grimme (G06)\cite{Grimme}  corrections to PBE.

\par In order to overcome the difficulties in density functional theory within standard functionals like the LDA or the GGA to calculate excited states,
 including the band gap which is underestimated, we have used the GW approximation
\cite{Hedin1,Hedin2} to obtain the bandstructure of solid nitromethane.
Instead of the Kohn-Sham equations used in DFT, the following quasiparticle equation is solved:
\begin{eqnarray}
(T+V_{ext}+V_{h})\psi_{{\bf k}n}({\bf r}) + \int d^3r^{\prime} \Sigma({\bf r},{\bf r}^{\prime},E_n({\bf k}))\psi_{{\bf k}n}({\bf r}^{\prime})\\= E_n({\bf k})\psi_{{\bf k}n}({\bf r}) \nonumber
\label{eq:qp_psi}
\end{eqnarray}
where $T$ is the kinetic energy operator, $V_{ext}$ is the external potential
 from the ion cores, $V_{h}$ is the Hartree potential, $\Sigma$ is the self-energy operator, and $E_n({\bf k})$ and $\psi_{{\bf k}n}({\bf r})$
are respectively the quasiparticle energy and the quasiparticle wave function.
Then, $\Sigma$
is approximated (vertex corrections are neglected) as the product of the Green's function $G$ times the screened Coulomb interaction $W$, which gives the GW approximation.
In practice, the GW quasiparticle eigenvalues are calculated as a correction to the DFT eigenvalues $\epsilon_n({\bf k})$ like\cite{Onida}:
%$\textrm{Re}E_n({\bf k})) = \epsilon^{DFT}_n({\bf k})+ Z_{n{\bf k}} \times$

\begin{eqnarray}\label{quasiparticle_energy_final}
\textrm{Re}~E_n({\bf k})& = & \epsilon^{DFT}_n({\bf k})+ Z_{n{\bf k}}\\ \nonumber
& & \times [\langle\Psi^{DFT}_{{\bf k}n}| \textrm{Re}\Sigma({\bf r},{\bf r}^{\prime},\epsilon_n({\bf k}))|\Psi^{DFT}_{{\bf k}n}\rangle \\ \nonumber
& & - \langle\Psi^{DFT}_{{\bf k}n}|V_{xc}^{DFT}(r)|\Psi^{DFT}_{{\bf k}n}\rangle]
\end{eqnarray}

with the QP renormalization factor $Z_{n{\bf k}}$ being:
\begin{equation}
Z^{DFT}_{n{\bf k}}=[1-\langle\Psi^{DFT}_{{\bf k}n}|
\frac{\partial}{\partial\omega} \textrm{Re}\Sigma({\bf r},{\bf r}^{\prime},
\epsilon_n({\bf k}))
|\Psi^{DFT}_{{\bf k}n}\rangle]^{-1}. \nonumber
\end{equation}

We have used the implementation of the G$_0$W$_0$ approximation\cite{seb1,seb2} provided by the code VASP(Vienna Ab-initio Simulation Package)\cite{Kresse}.
 To obtain convergence, we used 200 bands for the summation over the bands in the polarizability and the self-energy formulas, and the polarizability matrices were calculated up to a cut-off of $150$ eV.

\section{Results and discussions}
 \subsection{Structural properties}
As the starting point of our calculations, we adopted the experimental crystal structure of solid nitromethane\cite{Trevino} having the P2$_1$2$_1$2$_1$ space group and with lattice vectors
 of a = 5.183 \AA, b = 6.236 \AA, c = 8.518 {\AA}, which contains four molecules (28 atoms) of nitromethane per unit cell (see Fig. 1).
\begin{table*}[]
\caption{The calculated ground state properties of solid nitromethane at zero pressure. $a$, $b$ and $c$ are the lattice parameters, $V$ the volume of the orthorhombic unit cell. (in parenthesis: relative deviations from experiments (Ref. \onlinecite{Trevino}))
}
\begin{ruledtabular}

\begin{tabular}{ccccccc}
Method& XC       & a(\AA)   & b(\AA)   & c(\AA)   & V(\AA$^3$)    \\ \hline\\
   & LDA-CAPZ    & 4.91     & 6.03     & 8.10     & 239.82 \\
   &             & (-5.3\%) & (-3.3\%) & (-4.9\%) & (-12.9\%)\\
DFT& GGA-PW91    & 5.49     & 6.76     & 9.15     & 339.59 \\
   &             & (+5.9\%) & (+8.4\%) & (+7.4\%) & (+23.3\%)\\
 &GGA-PBE        & 5.47     & 6.73     & 9.10     & 334.98 \\
   &             & (+5.5\%) & (7.9\%)  & (+6.8\%) & (+21.7\%)\\\\
   \hline \\
 &GGA-PW91-OBS   & 5.05     & 6.24     & 8.44     & 266.43 \\
   &             & (-2.6\%) & (0.1\%)  & (-0.9\%) & (-3.2\%)\\
DFT-D&GGA-PBE-TS & 5.21     & 6.48     & 8.64     & 291.97 \\
   &             & (+0.5\%) & (3.9\%)  & (+1.4\%) & (+6.1\%)\\
 &GGA-PBE-G06    & 5.19     & 6.29     & 8.58     & 280.01 \\
   &             & (+0.1\%) & (+0.9\%) & (+0.7\%) & (+1.7\%) \\\\
% other works  & & 5.279$^a$, 5.317$^b$  & 6.569$^a$, 6.754$^b$ & 8.863$^a$, 8.980$^b$ & 302.27$^a$, 322.49$^b$\\
%        & &5.176$^c$, 5.165$^d$  & 6.274$^c$, 6.282$^d$ & 8.537$^c$, 8.526$^d$ & 277.30$^c$, 276.67$^d$\\\hline\\
   \hline \\
 Exp$^a$ &      & 5.183     & 6.235    & 8.518     & 275.31 \\
 \end{tabular}
\end{ruledtabular}
%$a:$ Ref. \onlinecite{Riad} (with GGA-PBE) ,\\  $b:$ Ref. \onlinecite{Liu} (with GGA-PW91),\\  $c:$ Ref. \onlinecite{conroy} (with PAW+PBE-vdW corrected),\\ $d:$ Ref. \onlinecite{dftd} (with GGA-G06),\\
$a:$ Ref. \onlinecite{Trevino} (Neutron diffraction at 4.2 K). \\
\end{table*}
The calculated equilibrium lattice parameters, volumes and relative deviations with experimental results are presented in Table I.
We observe a large difference between our calculated values obtained with either the LDA, GGA-PBE, or GGA-PW91 functional
(which are in reasonable agrement with earlier reported calculations\cite{Riad,Liu}) and the experimental datas. For instance,
the equilibrium volume obtained with LDA (239.8 \AA$^3$) is much lower than the experimental value (275.3 \AA$^3$), while the values obtained with GGA-PW91 (339.6 \AA$^3$) or GGA-PBE (334.98 \AA$^3$) are 20 \% too large.
However solid nitromethane is a molecular crystal where intermolecular interactions are dominated by van der Waals forces, and such
 interactions are not described correctly by standard functionals such as LDA or GGA.
To improve our results, we have used various DFT-D methods to describe vdW forces in our calculations (see table I), which show a
 good agrement with the experimental values.
Overall, the structural properties obtained with the PBE-G06 are showing the closest agrement with experiments.
 and therefore we will proceed further with this functional and also with the standard functionals LDA and PBE
  for comparison.
Notice that our results are in agreement with previous studies\cite{dftd, Landeville,Citroni} on solid nitromethane using similar methods.

\begin{figure}
\centering
\includegraphics[width=3.6in,clip]{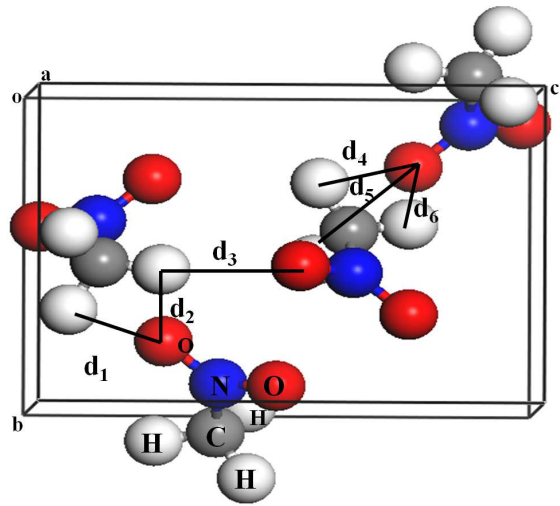}
\caption{(Colour online) Optimized crystal structure of solid nitromethane using GGA+G06 at 0 GPa. Here d$_{i}$ (i=1-6) are the different O-H bonds.}
\end{figure}

\begin{figure*}
\centering
\subfigure[]{\includegraphics[width=3.2in,clip]{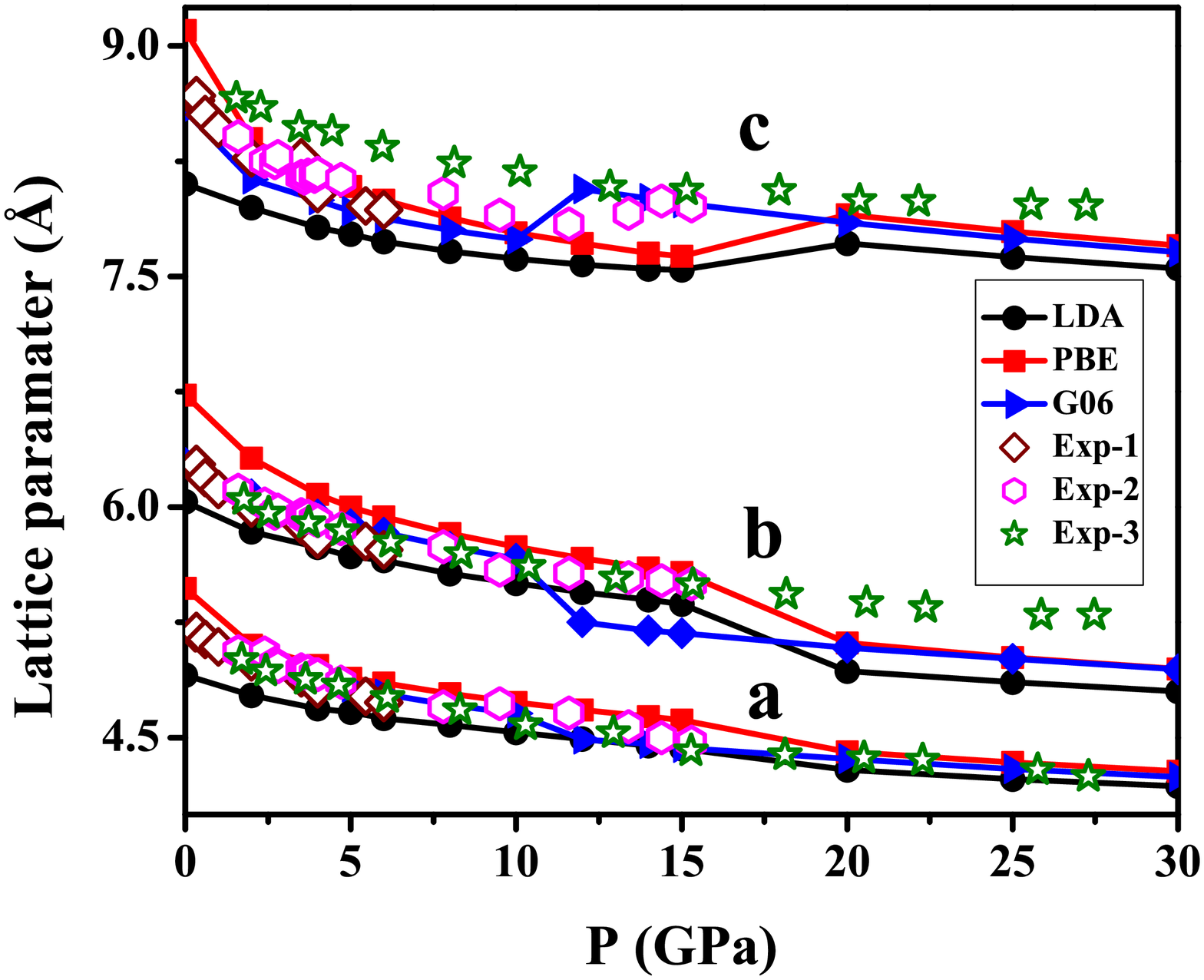}}
%\subfigure[]{\includegraphics[width=3.2in,clip]{NM-abc}}
\subfigure[]{\includegraphics[width=3.2in,clip]{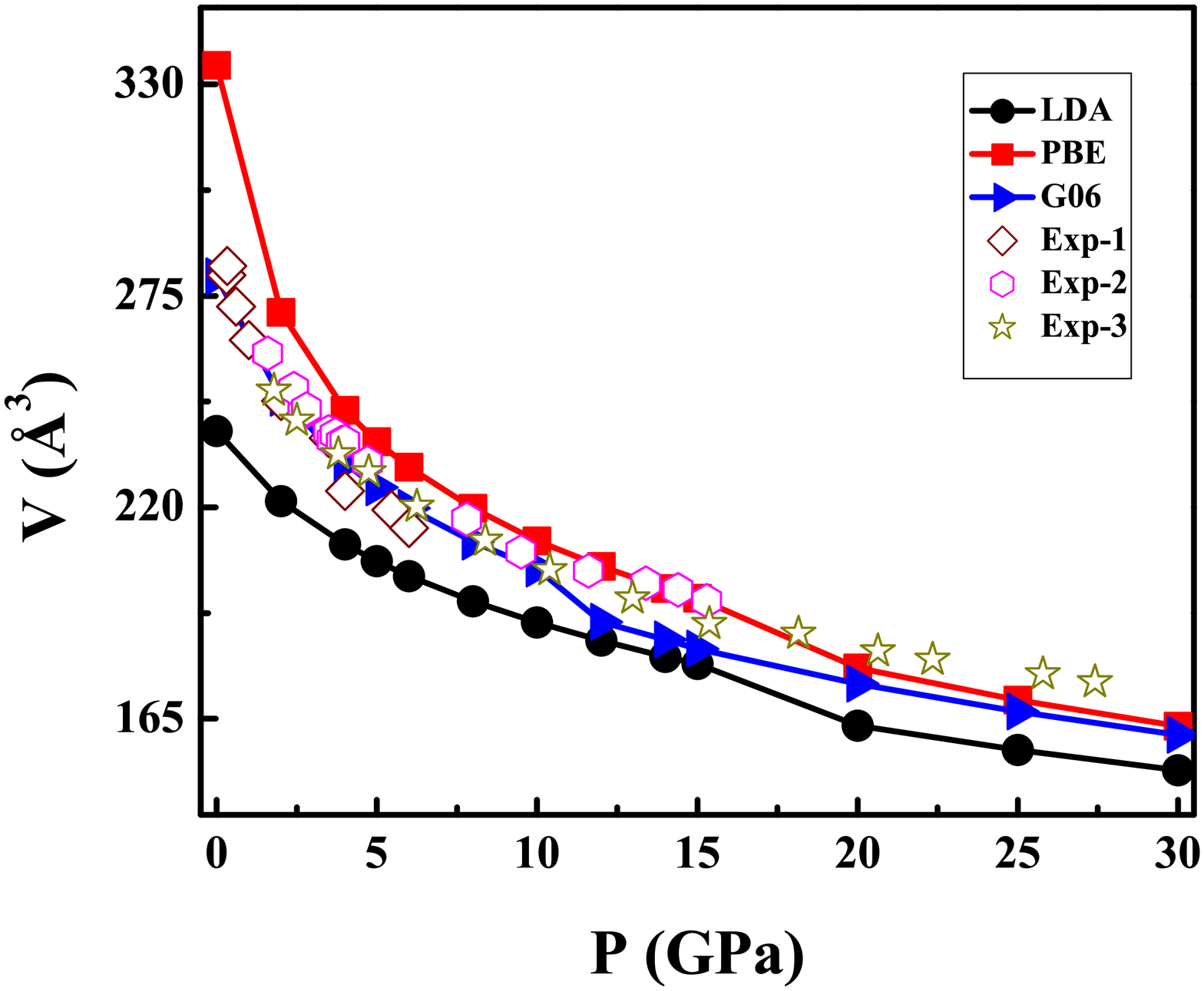}}
%\subfigure[]{\includegraphics[width=3.2in,clip]{NM-V}}
\caption{(Colour online) (a). Pressure dependence of lattice parameters a, b, and c (b). Comparison of the pressure dependence of the crystal volume of solid nitromethane up to 30 GPa as calculated within LDA, PBE, and PBE-G06, compared to experiments (Exp1: Ref. \onlinecite{Cromer}, Exp2: Ref. \onlinecite{yarger}:, Exp3: Ref. \onlinecite{Citroni})}.
\end{figure*}
\par In order to examine the effect of hydrostatic pressure on this material, we also computed the structural properties of solid nitromethane up to 30 GPa.
The variation in the lattice constants a, b, and c as function of pressure are compared with experiments in Fig. 2(a). To cover the various range of pressure
 of 0-6 GPa \cite{Cromer}, 0-15 GPa\cite{yarger} and 0-30 GPa\cite{Citroni} of the different experiments, we conduct our calculations from 0 GPa to 15 GPa with steps of 2 GPa and continue
 with steps of 5 GPa size up to 30 GPa. While at low pressure the +D correction to PBE is essential to reproduce the experimental values, at large pressure the three functionals
  give close results, which implies that at high pressures the long-range vdW forces become less important, as expected.
 Moreover it is seen that when using PBE-G06 functional, a sudden change in the lattice parameters occurs between 10 GPa to 12 GPa: the lattice parameters a and b
 decreases with increasing pressure whereas the c lattice parameter increases with pressure. A similar behavior is observed with the LDA and PBE functionals, although it happens
 at higher pressure (between 15 GPa to 20 GPa).

 To understand the sudden change in the lattice parameters, we have calculated
 the pressure dependence of the bond lengths and of the bond angles, the corresponding datas are shown in Fig. 3 and Fig. 4.
 The change observed in the lattice parameters is also reflected in the bond lengths and bond angles.
 In particular, when hydrostatic pressure is increased from 0 GPa to 10 GPa with the PBE-G06 functional, the bond lengths C-N, N-O$_i$ (i=1,2),
 O-H$_i$ (i=1,6) and the bond angle O$_1$-N-O$_2$  decreases significantly but in contrast, the C-N-O$_i$ (i=1,2) bond angles increases with pressure.
 However in the pressure region from 10 GPa to 12 GPa, an increase in the N-O$_2$ bond length is observed together with a sudden change in the C-N-O bond angles.
 Also, the O-H distances (see Fig. 4) d$_1$, d$_3$ and d$_6$ are in the range of 2.2~\AA~ to 3.2~\AA~ at zero pressure, whereas these values decreases to 2.0~\AA~ to 2.6~\AA~ at high pressures and we found weakening of hydrogen bond strength in 10 to 12 GPa.
 This behavior indicates slight changes in the hydrogen bonding and more generally in the intermolecular geometry of solid nitromethane.
 In Fig. 5. we have pictured the unit cell of solid nitromethane for different pressures. A phase transition is clearly observable,
 which corresponds to the Phase I to Phase II known experimentally\cite{Citroni} to occur at 11 GPa.
  In our case the transition pressure is slightly different, but the overall agreement is found to be good.

\begin{figure*}
\centering
\includegraphics[width=6.5in,clip]{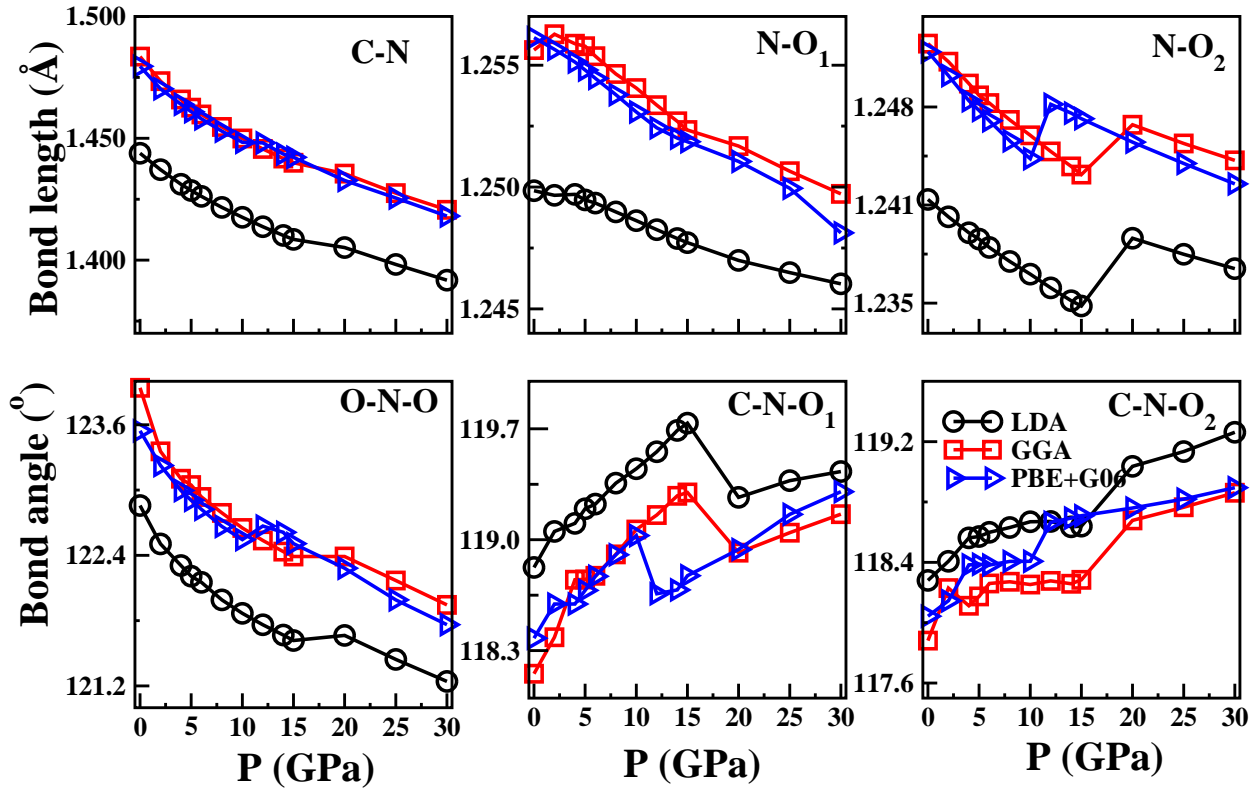}\\
\caption{(Color online) Bond lengths and bond angles of solid nitromethane up to 30 GPa as calculated within LDA, PBE, and PBE-G06.}
\end{figure*}

\begin{figure}
\centering
\includegraphics[width=3.2in,clip]{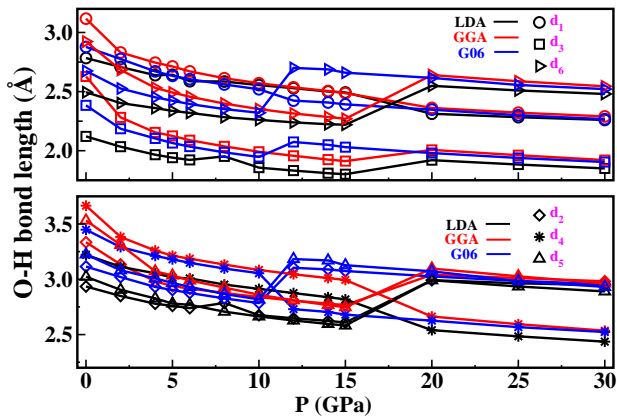}\\
\caption{(Color online) Oxygen-hydrogen bond lengths of solid nitromethane up to 30 GPa as calculated within LDA, PBE, and PBE-G06.}
\end{figure}

\begin{figure*}
\centering
\includegraphics[width=7in,clip]{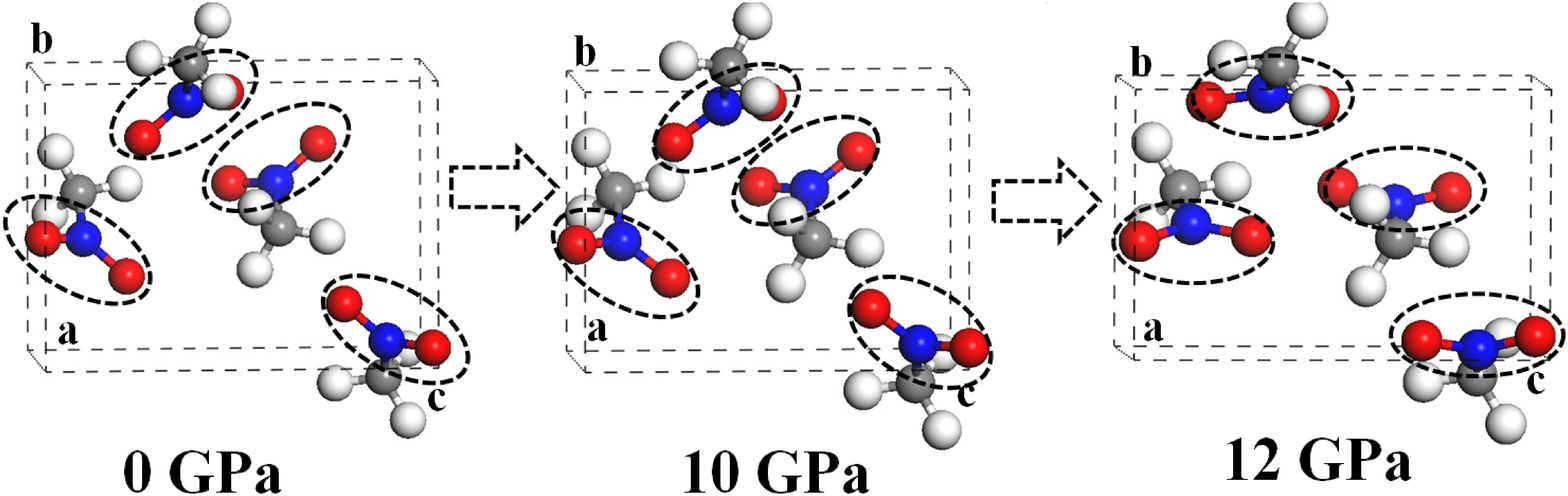}
\caption{(Color online) Snap shots of solid nitromethane unit cells using PBE+G06 at 0 GPa, 10 GPa and 12 GPa. The circles shows the orientation of C-N-O bond angle with pressure.}
\end{figure*}

\par Fig 2(b) shows our calculated pressure-volume relation up to 30 GPa and compared with experiments.
%As mentioned earlier, the equilibrium volumes
% using LDA and PBE functional are largely different from experiments, the bulk moduli calculated with these approaches are significantly alter than experiments.
 The calculated bulk modulus and its pressure derivative obtained with the  PBE-G06 functional are B = 11.6 GPa and B' = 6.5 respectively, which is in good agreement with
 experiments 7  GPa,\cite{Cromer} 10.1 GPa,\cite{yarger} 9.25($\pm$1.9) GPa\cite{Citroni}  and closely comparable with the reported theoretical values 5.7 GPa,\cite{Sorescu} 5.7 GPa.\cite{Liu}

 \subsection{Elastic properties}

 To have a better understanding of the mechanical properties of nitromethane, we have studied the corresponding elastic properties.
  We have used the PBE-G06 functional since it provided us with lattice parameters in close agreement with experiments (See Table I).
 Solid nitromethane crystalizes in an orthorhombic structure, it has 9 independent elastic constants
 namely C$_{11}$, C$_{22}$, C$_{33}$,  C$_{44}$, C$_{55}$,  C$_{66}$, C$_{12}$, C$_{13}$ and C$_{23}$. To calculate the elastic properties, we have used the volume-conserving
 strain technique\cite{Mehl} and the calculated single crystal elastic constants are presented in Table II.

\begin{table}[ht]
\caption{
Single crystal elastic constants (C$_{ij}$, in GPa)  of solid nitromethane. All quantities are calculated at the
respective theoretical equilibrium volume obtained with the GGA-G06 functional.}
\begin{ruledtabular}
\begin{tabular}{cccccccccccccccc}
C$_{11}$ & C$_{22}$ & C$_{33}$ & C$_{44}$ & C$_{55}$& C$_{66}$ & C$_{12}$ & C$_{13}$ & C$_{23}$ &\\\hline
20.3 & 20.1 & 16.1 & 5.2 & 1.4 &6.2&7.4 & 7.2 & 4.6 &
\end{tabular}
\end{ruledtabular}
%$a$ Ref. \onlinecite{Cromer}, $b$ Ref. \onlinecite{yarger}, $c$ Ref. \onlinecite{Citroni}
\end{table}

The calculated values satisfy the Born stability criteria\cite{Born} for an orthorhombic structure and indicates that the system is mechanically stable.
 For an orthorhombic structure, the elastic constants C$_{11}$, C$_{22}$, and C$_{33}$ are directly related to the compression of crystallographic axes
 along the a, b, and c axes. In the case of solid nitromethane, they follow an order of C$_{11}$$>$ C$_{22}$ $>$ C$_{33}$ which implies that C$_{11}$ is the stiffest elastic constant:
  one can expect  strong interactions along the a-axis, whereas it has more sensitive to detonation along the c-axis.
 A similar conclusion was obtained for solid nitromethane by Conroy et al\cite{conroy} by applying uniaxial
 compressions along different crystallographic planes.
 The other diagonal elements of the elastic tensors, C$_{44}$, C$_{55}$, and C$_{66}$, indicate the
 shear elasticity applied to two dimensional regular lattices along [1 0 0], [0 1 0] and [0 0 1].
 From our calculations,  we found that C$_{66}$$>$ C$_{44}$$>$  C$_{55}$, which indicates a softer shear transformation along the [1 0 0] and [0 1 0] directions in solid nitromethane
  than along the [0 0 1] direction. The off diagonal elastic components are found to be the order of C$_{12}$$\geq$ C$_{13}$$>$ C$_{23}$. This implies that C$_{12}$ and C$_{13}$ couple an
 applied normal stress component in the crystallographic axis a direction with uniaxial stress along b and c axes. Though there is no experimental elastic constants available for this material,
 at present we compared the calculated elastic properties with other secondary explosive materials RDX,\cite{RDX} HMX\cite{HMX} and PETN.\cite{PETN}
 Interestingly in all these explosives, C$_{11}$$, $ C$_{22}$ and C$_{33}$ follow the same ordering of C$_{11}$$>$ C$_{22}$ $>$ C$_{33}$ and this implies
 that most of the secondary explosives have more sensitive along crystallographic c-axis and less sensitive along a-axis.
 % which similar in other explosives RDX\cite{RDX}, HMX\cite{HMX} and PETN\cite{PETN}.
 Finally the calculated single crystal bulk modulus from elastic constants is 10.32 GPa, which is excellent agreement with earlier experiments.\cite{Cromer,yarger,Citroni}
 From this, we can say that the calculated elastic constants and bulk modulus are quite reasonable.

\subsection{Vibrational properties}

In this section, we report on our investigation of the vibrational properties of solid nitromethane using linear response method with in density functional perturbation theory,
 as implemented in the CASTEP code. Since the unit cell of solid nitromethane contains 28 atoms, the corresponding number of modes are 84, of which 3 are acoustic modes and remaining 81 modes are optical modes.  The optical modes at the Brilloun zone center ($\Gamma$ point) have the following irreducible representation: A(R),  B$_1$(R+IR), B$_2$(R+IR) and B$_3$(R+IR). Here R and IR represents the Raman active and infrared active respectively.
 Table III and IV provides the calculated internal and external modes for PBE+G06 calculations, with the corresponding mode assignment and compared with earlier experimental  and other theoretical results. By following
 the previous notation, the high energy domain($>$ 400 cm$^{-1}$) contains 14 internal modes where each mode  grouped into four set of symmetries. These modes are indexed going from
 highest energy mode M1 (CH$_3$ antisymmetric stretching) to the lowest energy mode M14 (NO$_2$ rocking). The main features of the internal modes are as follows (1) The higher
 frequency modes M1, M2 and M3 are mainly from CH$_3$ stretching, (2) optical modes M5 to M8 are because of CH$_3$ deformation and NO$_2$ stretching, (3) NCH deformation is observed
 in mode M9, (4) mixed motions are observed in M10 to M13, which are from NO$_2$, CH$_3$ and CN bonds (5) M14 mode is mainly from the NO$_2$ rocking. The calculated internal modes differ
 by few cm$^{-1}$ by comparing with earlier reported GGA-PBE calculated frequencies.\cite{Liu} This is mainly from due to the inclusion of vdW correction to GGA which imporoved the volume
 compared by without inclusion of vdW and thereby intermolecular distance varies considerably. Whereas the low energy domain($<$ 400 cm$^{-1}$) consists 25 fundamental external (lattice)
 modes, namely 7A+6B$_1$+6B$_2$+6B$_3$. The internal and external modes are separated by 292.6 cm$^{-1}$, which is in good agrement with experimental gap of 289.2 cm$^{-1}$.
 %The difference between the external and internal modes is 292.6 cm$^{-1}$, which is in good agrement with experiment gap 289.2 cm$^{-1}$. This gap
% indicates that the molecules behave like separated rigid bodies for lower and higher energy domains.
 Overall, the calculated external and internal modes are remarkably well in agrement
 with experiments.\cite{Ouillon, OOuillon, Pinan}

\begin{table*}[]
\centering
\caption{Vibrational frequencies (in cm$^{-1}$) of the internal modes of solid nitromethane. All quantities are calculated at the
respective theoretical equilibrium volume using the GGA+G06 functional. Here A(R), B$_1$(R+IR), B$_2$(R+IR) and B$_3$(R+IR) are irreducible representation of space group P2$_1$2$_1$2$_1$.}
%\begin{longtable}{ccccccccccccccccccccccccc}
\begin{ruledtabular}
\begin{tabular}{ccccccccccccccccccccccccc}
no. & Mode &Irr rep & Cal. fre & Ass. & Exp\cite{Ouillon} & other (PBE)\cite{Liu} \\ \hline
1 & & B$_1$ & 3127.8 & \\
2 &M1 &B$_2$ & 3127.3 & CH$_3$ antisymmetric stretching & 3082.2 (B$_2$) & 3117\\
3 & &B$_3$ & 3126.5 & \\
4 & &A     & 3126.3 &\\

5 &  &B$_2$ & 3082.7 & \\
6 & M2 &B$_1$ & 3081.6 & CH$_3$ antisymmetric stretching & 3049.6 (B$_3$) & 3072 \\
7 &  &B$_3$ & 3081.2 & \\
8 &  &A     & 3080.9 & \\

9 &  &B$_2$ & 2983.1 & \\
10&M3 &B$_1$ & 2981.8&CH$_3$ symmetric stretching   & &2980 \\
11& &B$_3$ & 2981.4 & \\
12&  &A     & 2981.0 && 2970.6 (A) \\

13 & &B$_3$ & 1483.1 & & 1566.0 (B$_3$)\\
14 & &B$_1$ & 1480.1 & CH$_3$ wagging +  \\
15 &M4 &B$_2$ & 1479.2 & NO$_2$ antisymmetric stretching & 1565.4 (B$_2$) & 1528 \\
16 & &A     & 1456.3 & \\

17 & &B$_1$ & 1434.5 &  \\
18 &M5 &B$_3$ & 1429.3 & CH$_3$ deformation +  &1429.6 (B$_3$) \\
19 & &B$_2$ & 1419.1 & NO$_2$ antisymmetric stretching & 1430.9 (B$_2$) & 1434\\
20 & &A     & 1418.6 & \\

21 & &B$_2$ & 1406.9 &  & 1414.1 (B$_2$)\\
22 & &B$_1$ & 1396.6 & CH$_3$ deformation + &1412.7 (B$_1$) \\
23 &M6 &B$_3$ & 1384.1 & NO$_2$ symmetric stretching & 1413.6  (B$_3$) & 1414 \\
24 & &A     & 1381.3 & & 1412.0 (A) \\

25 & &B$_2$ & 1371.4 &  & 1414.7 (B$_2$) \\
26 &M7 &B$_1$ & 1368.2 &CH$_3$ deformation + \\
27 & &B$_3$ & 1360.6 &  NO$_2$ symmetric stretching &  & 1387  \\
28 & &A     & 1356.0 && 1403.9 (A)\\

29 & &B$_2$ & 1310.3 &  & 1376.5 (B$_2$)\\
30 &M8 &B$_3$ & 1310.1 & CH$_3$ wagging + NO$_2$ stretching+ &1378.2 (B$_3$) &1342 \\
31 & &B$_1$ & 1308.2 & CN stretching & 1379.3 (B$_1$)\\
32 & &A     & 1307.9 && 1375.4 (A)\\

33 & &B$_2$ & 1095.4 && 1120.1 (B$_2$) \\
34 &M9 &B$_1$ & 1090.5 &  NCH deformation & 1121.2 (B$_1$) &1098  \\
35 & &B$_3$ & 1089.5 && 1124.3 (B$_3$)  \\
36 & &A     & 1088.3 & \\

37 & &B$_3$ & 1076.7 &&1108.2 (B$_3$) \\
38 & &A     & 1075.5 & CH$_3$ twisting + & 1105.3 (A)\\
39 &M10 &B$_2$ & 1073.5 & NO$_2$ antisymmetric stretching & 1107.1 (B$_2$)  & 1083 \\
40 & &B$_1$ & 1070.7 &&1106.1 (B$_1$)  \\

41 & &A     & 890.1 & &923.6 (A) \\
42 & &B$_2$ & 889.9 & CN stretching +  & 923.8 (B$_2$) & 906 \\
43 &M11 &B$_1$ & 889.1 & NO$_2$ bending & 922.0 (B$_1$)  \\
44 & &B$_3$ & 888.9 &  \\

45 & &B$_1$ & 641.5 && 658.0 (B$_1$) \\
46 &M12 &B$_2$ & 641.1 & CN stretching +  & 664.4 (B$_2$) & 650\\
47 & &B$_3$ & 641.0 & NO$_2$ scissor \\
48 & &A     & 636.3 & \\

49 & &B$_1$ & 592.4 &&  610.1(B$_1$)\\
50 & M13 &B$_2$ & 591.4 & NCH wagging + & 608.0 (B$_2$)\\
51 & &B$_3$ & 587.3 & NO$_2$ antisymmetric stretching & 607.7 (B$_3$) & 601 \\
52 & &A     & 585.9 & \\

53 & &B$_2$ & 472.7 &&484.1 (B$_2$)   \\
54 &M14 &A  & 470.9 &  & 484.9 (A) &472 \\
55 & &B$_3$ & 470.6 & NO$_2$ rocking & 484.0 (B$_3$)\\
56 & &B$_1$ & 470.3 && 483.7 (B$_1$)

%\end{longtable}
\end{tabular}
\end{ruledtabular}
\end{table*}

\begin{table*}[]
\caption{
Vibrational frequencies (in cm$^{-1}$) of the external modes of solid nitromethane. All quantities are calculated at the
respective theoretical equilibrium volume using the GGA+G06 functional. Here A(R), B$_1$(R+IR), B$_2$(R+IR) and B$_3$(R+IR) are irreducible representation of space group P2$_1$2$_1$2$_1$.  }
\begin{ruledtabular}
\begin{tabular}{cccccccccccccccc}
no.& Mode & Irr rep & Cal. fre &Exp\cite{Ouillon}\\ \hline
57 && A     & 180.9 & &\\
58 && B$_2$ & 171.1 & 165.4(B$_2$)&\\
59 && B$_1$ & 162.0 & \\
60 &L1& B$_3$ & 161.4 &  \\
61 && A & 153.6 & 154.9(A)\\
62 && B$_2$ & 148.5 & \\
63 && B$_3$ & 145.7 & \\

64 &L2& B$_1$     & 132.9 &  138.8(B$_1$)\\

65 && B$_3$ & 116.1 & 114.6(B$_3$) \\
66 && B$_1$ & 113.0 & \\
67 &L3& B$_2$ & 107.7 & \\
68 && A     & 106.3 & 112.9(A)\\
69 && B$_3$ & 100.5 & \\

70 && B$_2$ & 97.6 & 95.1 (B$_2$)\\
71 &L4& B$_1$ & 91.2 & 99.7(B$_1$)\\
72 && B$_3$ & 90.5 &85.0(B$_3$) \\
73 && A     & 84.4 & 84.5(A) \\

74 && B$_1$ & 73.1 &78.2(B$_1$) \\
75 && B$_2$ & 71.6 &71.5(B$_2$) \\
76 &L5& A     & 70.1 &70.7(A)&   \\
77 && A     & 66.1 & 58.6(A) \\

78 && B$_3$ & 62.7 & \\
79 && A     & 53.7 &52.1(A) \\
80 &L6& B$_2$ & 51.8 &44.4(B$_2$)&  \\
81 && B$_1$ & 44.1 &

\end{tabular}
\end{ruledtabular}
\end{table*}

\par Now, we discuss the pressure dependence of the vibrational frequencies of solid nitromethane. Fig. 6 shows the frequency of the internal modes in function of
 pressure up to 30 GPa (with a step size of 5 GPa). It is observed that the frequencies of all the internal modes increases with external pressure. However,
 in the pressure range 10 GPa to 15 GPa, a large discontinuity is noticed for the vibrational modes M2, M3, M4, M7 and M8. From Table. III, it is found that the mentioned
 modes are from CH$_3$ and NO$_2$ stretching modes at ambient conditions and hydrostatic compression shows significant effect on CH$_3$ and NO$_2$ groups.
 A similar behavior has been observed from 9 to 13 GPa in the  experiments using Raman measurements\cite{Ouillon,Pinan,OOuillon}.
 The variation under pressure of lattice modes are shown in Fig. 7. From this, the calculated lattice modes from L1 to L3 (Fig. 7(a)) are harding with increasing pressure, whereas L4 to L6 modes(Fig. 7(b)) starts softening from 5 GPa to 10 GPa. As the pressure increases from 10 GPa, all these modes(L4 to L6) shift towards higher frequencies. However in the pressure region 8 to 10 GPa, B$_1$ modes decreases to zero and shown in Fig. 7(c).
 Overall, the calculated vibrational frequencies at high pressure using PBE-G06 functional shows
 distinct intra and intermolecular modes at high pressures and the discontinuity in the internal and lattice  modes point out to the structural transition  in solid nitromethane between 10 to 12 GPa at low temperatures which is in good agreement with available experimental data.\cite{Citroni, Pinan}
 From 15 to 30 GPa, all the internal modes increases continuously and without any slope change, in agreement with earlier experiments. Also,
 by increasing the pressure the energy gap between the internal and the external modes decreases to
 30 cm$^{-1}$, which is comparable with previous experiment.
 \begin{figure*}
\centering
\includegraphics[width=6in,clip]{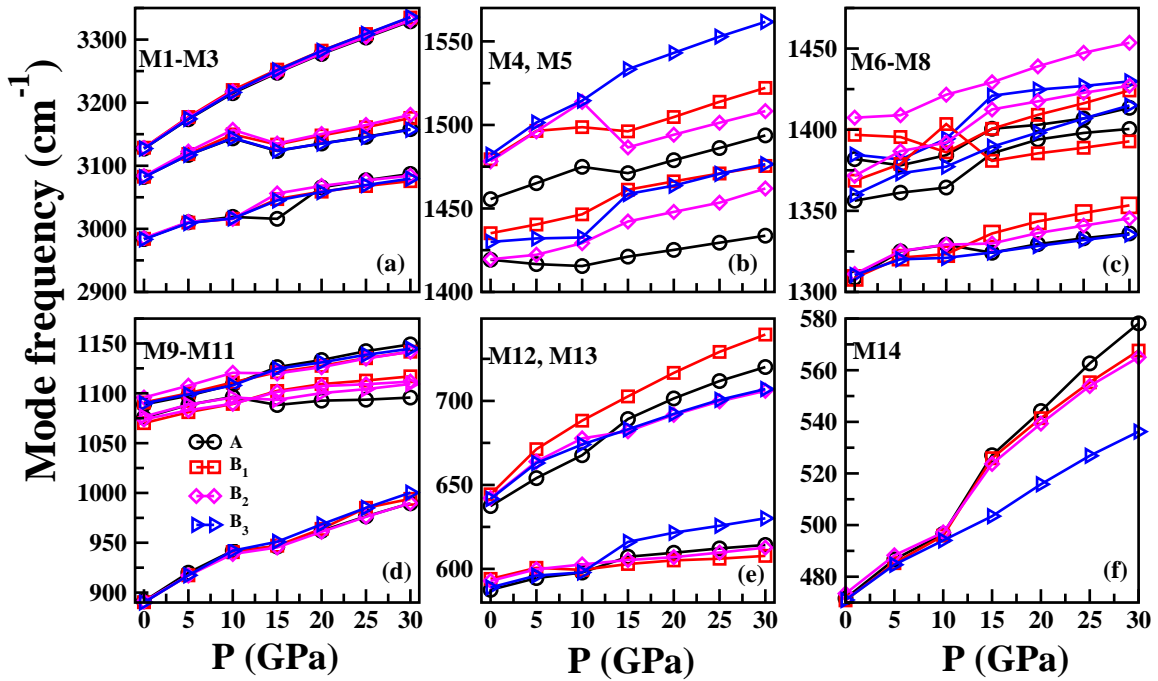}\\
\caption{(Colour online) Pressure evolution of internal vibrational frequencies of solid nitromethane up to 30 GPa using PBE-G06 at Gamma point. }
\end{figure*}

 \begin{figure*}
\centering
\includegraphics[width=6in,clip]{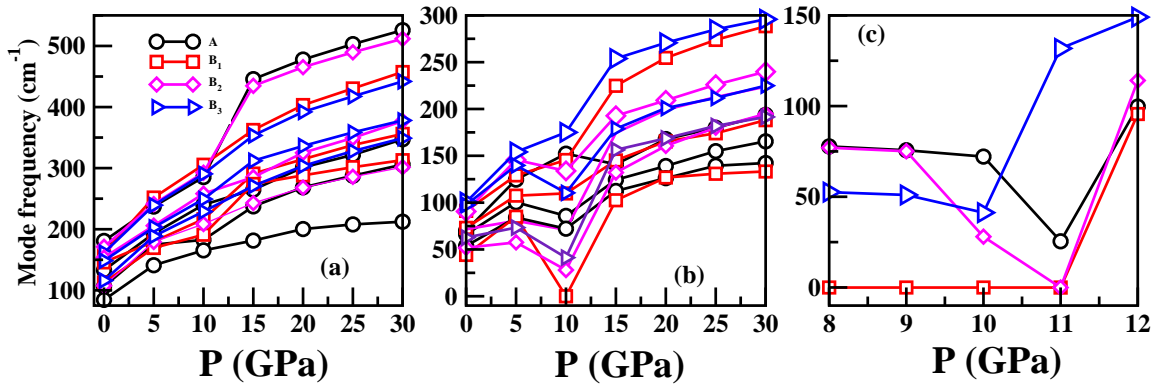}\\
\caption{(Colour online) Calculated L1-L3 (a), L4-L6 (b) external vibrational frequencies of solid nitromethane at Gamma point up to 30 GPa (step size of 5 GPa) within PBE-G06 and (c) shows the L6 mode frequencies from 8 GPa to 12 GPa with a step size of 1 GPa.  }
\end{figure*}

\subsection{Electronic properties}
As mentioned earlier, in order to correlate the impact sensitivity of an explosive material with its electronic band gap, it is necessary to know the electronic
band structure and the value of the band gap precisely. In particular, Wei and Xiao \cite{wei} correlated the impact sensitivity with the band gap of azides and confirmed that
 the materials having the smallest the band gap are easier to decompose and to explode under external stimuli.
 In the case of nitromethane, there is no experimental evidence for the value of band gap to the best of our knowledge.
Our calculated bandstructure along high symmetry directions obtained with the GGA approximation is presented in Fig 8 (solid lines).
The top of the valence band and the bottom of the conduction band occur at the $\Gamma$ and X high symmetry points respectively, and the magnitude
 of our GGA bandgap (3.8 eV) is in good agrement with the previously reported theoretical values of 3.61 eV \cite{Liu} and of 3.28 eV.\cite{Manaa}
 However band-gaps obtained using standard DFT functionals are usually underestimated when compared to experiments. To solve this problem, we have used the GW approximation to treat excited states
 as implemented in the VASP code using projector augmented waves method.\cite{PAW}
The calculated quasiparticle band structure of solid nitromethane is presented in Fig. 8 (red circles), where it is compared to its GGA counterpart (solid lines).
\begin{figure}
\centering
\includegraphics[width=2.7in,clip,angle=270]{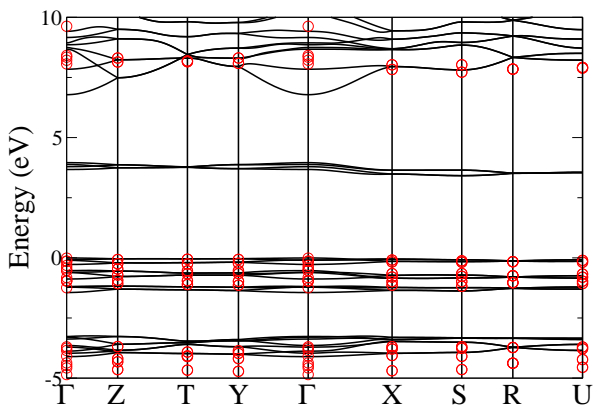}\\
\caption{(Color online) Band structure of solid nitromethane within GGA (black lines) and G$_0$W$_0$ approximation (red circles) at experimental lattice vectors.}
\end{figure}
Our calculated GW band gap value is 7.8 eV, which is a significant correction of the GGA value. As with the GGA, the band-gap is indirect
 from the $\Gamma$ point to the X point.
 From the calculated band structure as shown in Fig 8, the valence band splits into four distinct manifolds. The lower manifold that extends from about -3 eV to -4 eV is derived from
 the $p$ states of C atoms and partly of $s$ states of H atoms. The upper manifold that extends up to the Fermi level is mostly from the hybridization of $p$ states of C and O atoms.
 respectively. The middle regions,  that are  -0.5 eV to -1.0 eV and -1.5 to -2 eV are due to the hybridization of C and O atoms. The conduction band is mainly
 dominated by the H-s states and from states of O and N atoms.
 Moreover, the important correction brought by the GW approximation in the case of nitromethane implies that the GGA values of the band gap usually used for
  other secondary explosives to predict their exact sensitive impact factor needs to be reinvestigated.

\subsection{Optical properties}

 \begin{figure}[]
\begin{center}
{\subfigure[]{\includegraphics[height = 2.15in, width=3.0in]{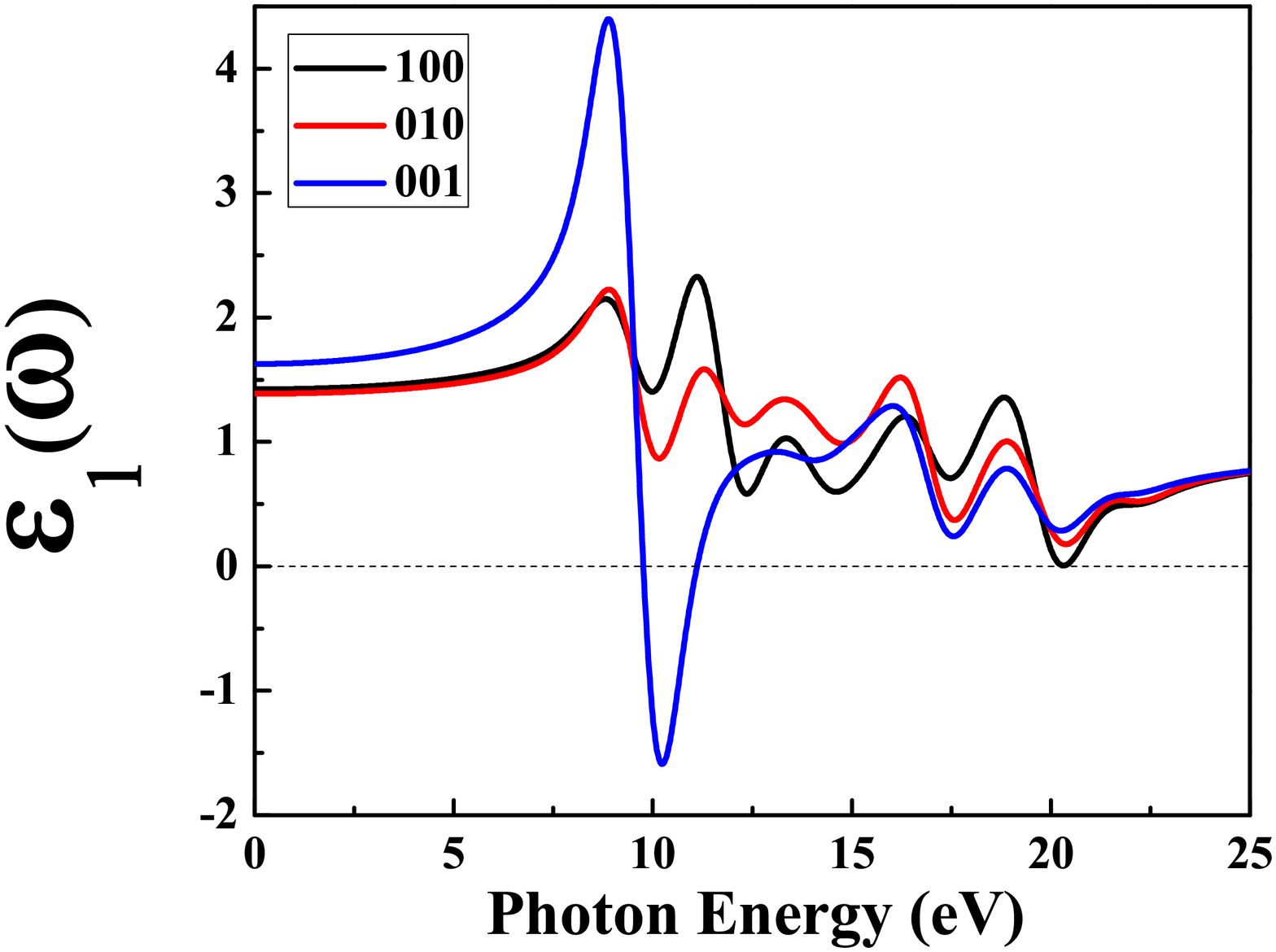}}}
{\subfigure[]{\includegraphics[height = 2.15in, width=3.0in]{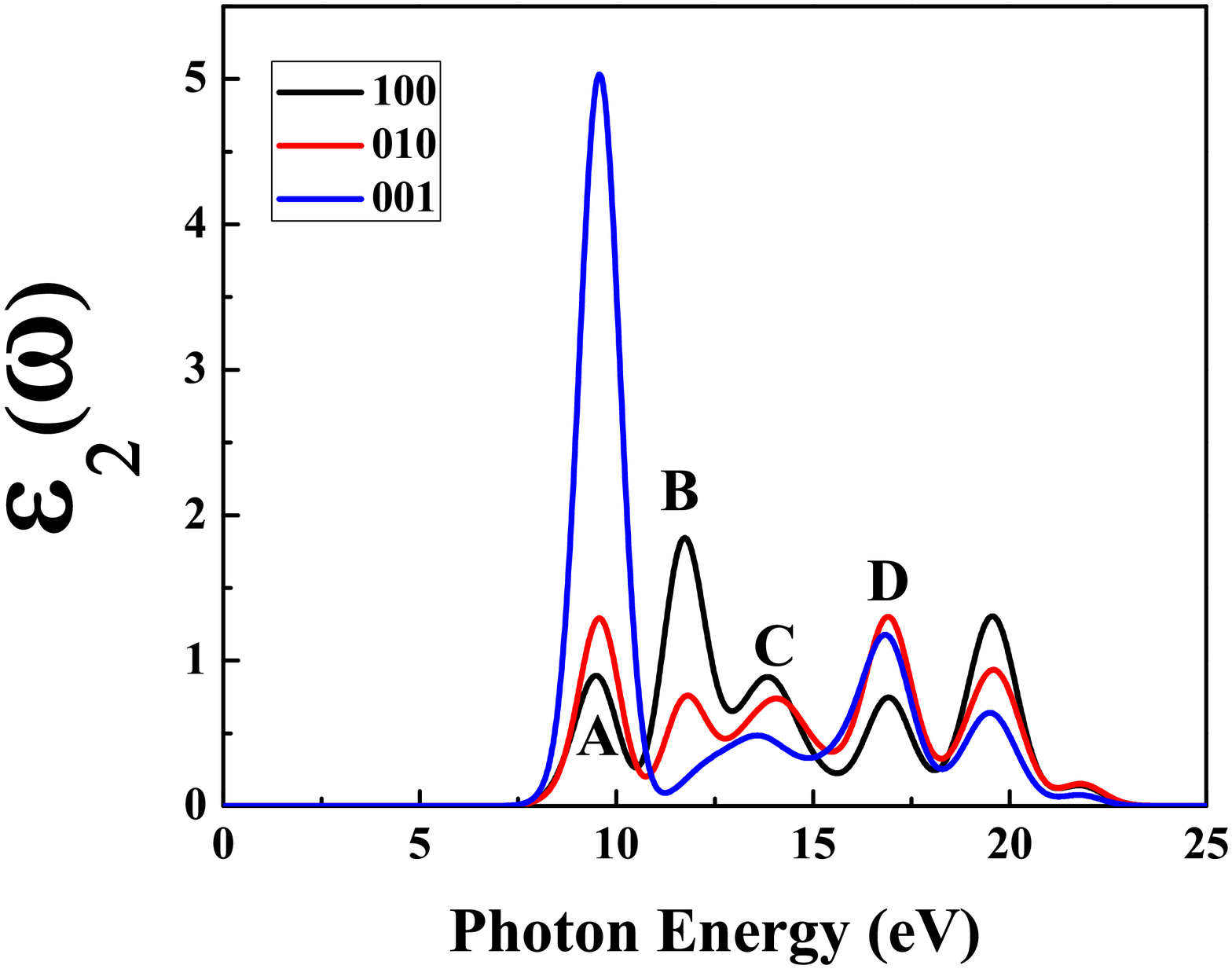}}}
{\subfigure[]{\includegraphics[height = 2.15in, width=3.0in]{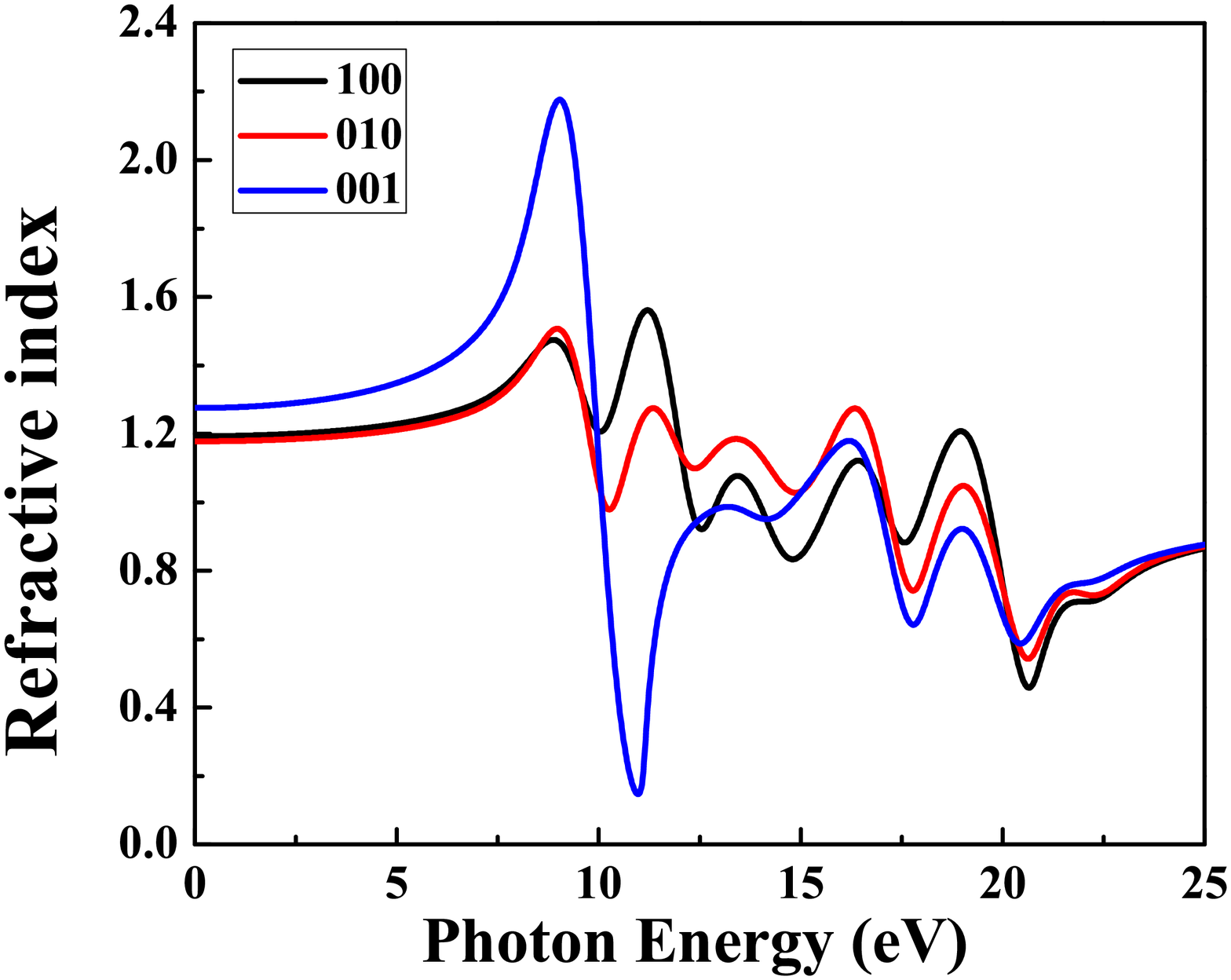}}}
{\subfigure[]{\includegraphics[height = 2.15in, width=3.0in]{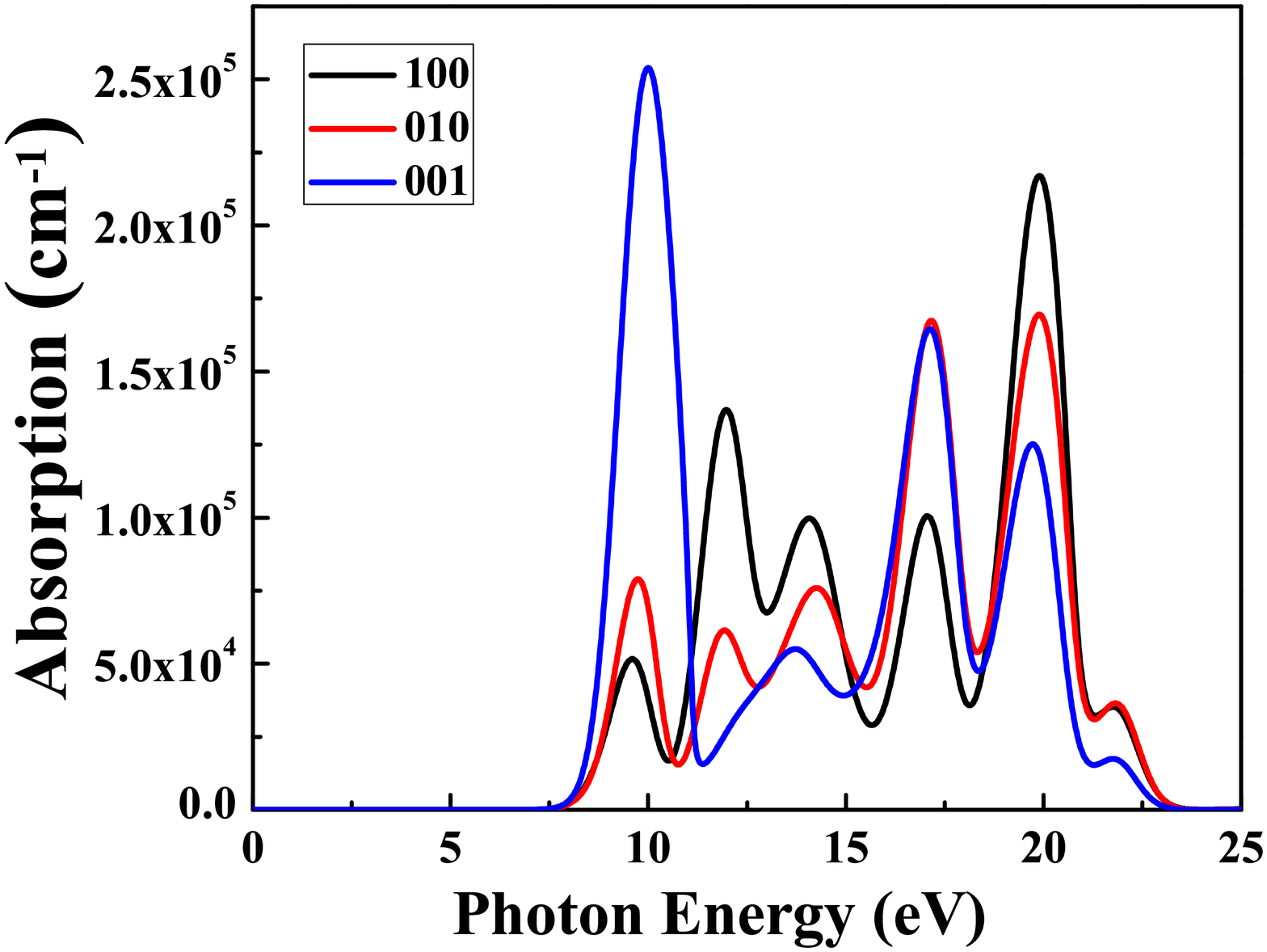}}}
%{\subfigure[]{\includegraphics[height = 2.15in, width=3.0in]{e1}}}
%{\subfigure[]{\includegraphics[height = 2.15in, width=3.0in]{e2}}}
%{\subfigure[]{\includegraphics[height = 2.15in, width=3.0in]{n}}}
%{\subfigure[]{\includegraphics[height = 2.15in, width=3.0in]{ab}}}
\caption{(Color online) Optical properties of solid nitromethane calculated at theoretical equilibrium with a scissor shift of 4.6 eV. }
\end{center}
\end{figure}
In this section, we focus on the optical properties of solid nitromethane.
 In general,  the optical properties of matter can be described by means of the complex
 dielectric function $\epsilon (\omega)$ = $\epsilon_1 (\omega)$ + $i\epsilon_2 (\omega)$, where $\epsilon_1 (\omega)$ and $\epsilon_2 (\omega)$ describes the
 dispersive and absorptive parts of the dielectric function. Generally,  there are two contributions to $\varepsilon$($\omega$), namely the intraband and interband
 parts. The contribution of intraband transitions are important only for metals while the interband transitions can be split into direct and indirect transitions.
 We have neglected the indirect transitions involving scattering of phonons assuming that they give small contribution to $\varepsilon$($\omega$) in comparison to
 direct transitions.\cite{smith} Then $\varepsilon$(\emph{q=0}, $\omega$) is calculated using the random phase approximation without local field effects. In
  particular, the imaginary part is determined by a summation over empty states, and then the real part $\epsilon_1 (\omega)$ of the dielectric function can be evaluated
 from $\epsilon_2 (\omega)$ using Kramer-Kroning relations. The knowledge of the real and imaginary parts of the dielectric function allows the
 calculation of the important optical properties such as refractive index and absorption. Since, solid nitromethane crystalizes in an orthorhombic structure,
 the optical functions are expected to be different along the three crystallographic directions [1 0 0], [0 1 0] and [0 0 1].
  We have calculated the optical properties using the CASTEP code within GGA but a scissor shift of 4.6 eV is applied to the conduction
  bands to be consistent with the calculated value of the band gap obtained with the GW approximation.
\par Our calculated optical functions of solid nitromethane are presented in Figure 9 for an energy range up to 25 eV.
  The dispersive part of the dielectric function ($\epsilon_1 (\omega)$) is shown in Fig 9 (a). The static dielectric function $\epsilon_1 (0)$ of
  solid nitromethane along the three directions are found to be $\epsilon_1 (0)$= 1.42 along the [1 0 0] direction, $\epsilon_1 (0)$= 1.39 along the [0 1 0] direction and $\epsilon_1 (0)$=1.63
  along the [0 0 1] direction.
 The absorptive part of the dielectric function ($\epsilon_2 (\omega)$) is shown in Fig. 9(b). The main peaks in the [1 0 0] direction are marked as A (at 9.5 eV),
  B (at 11.7 eV), C (at 13.8 eV) and D (at 16.9 eV). A similar structure is observed in the case of [0 1 0] and [0 0 1] crystallographic directions albeit with
  a slight energy shift and different intensities.
  The peak A originates from transitions between O (p) and H (s) states for all the three crystallographic directions.
  The peak B arises from transitions of the C (p) states to the H (s) states along [1 0 0] and [0 1 0] directions, while it is absent for the [0 0 1] direction. The peak
  C along the three directions comes from transitions between the s states of O and the p states of C to the p states of N and O atoms.
  The peak D results of transition from C, N, O p states to O, N 2s states along the
  three crystallographic directions. Then from the calculated $\epsilon_1 (\omega)$ and $\epsilon_2 (\omega)$, we have evaluated the refractive index (n) and absorption
  shown in Fig. 9(c) and Fig. 9(d) respectively. The static refractive index n(0) along three directions: n(0) = 1.19 along [1 0 0], n(0) = 1.18 along [0 1 0] and  n(0) = 1.27 along [0 0 1].
  In Fig 7(d), we show the calculated absorptive spectra along the three directions . The absorption
  starts at an energy of 7.8 eV which is the band gap, whereas the edge of absorption occurs at 9.60 eV along the  [1 0 0] direction, at 9.77 eV along the [0 1 0] direction,
  and at 9.98 eV along the [0 0 1] direction. The related absorptive coefficients are found to be different along  the three directions and are around $5 \times10^{4} cm^{-1}$ (along [1 0 0]),
  $7.8 \times10^{4} cm^{-1}$ (along [0 1 0]) and $2.5 \times10^{5} cm^{-1}$ (along [0 0 1]).
  From these results, it appears that solid nitromethane is an optically anisotropic material.

 \section{Conclusions}
In summary, we have investigated the structural properties of solid nitromethane within LDA and GGA with different functional including dispersion corrections to treat vdW forces.
It was found that the structural properties using PBE-G06 are in good agreement with experiments, which highlight the role of vdW interactions in solid nitromethane.
Then, the influence of hydrostatic pressure on the
structural properties such as the lattice parameters, the bond lengths, and the bond angles was calculated and
 a discontinuity in bond lengths and bond angles was observed between 10 GPa to 12 GPa.
Also, we have estimated the elastic constants for this material using PBE-G06 and found that the material is mechanically stable.
In particular, solid nitromethane is stiffest along the crystallographic a-axis, followed by the b-axis, and then by the c-axis.
The vibrational properties of solid nitromethane at ambient conditions have been calculated and a fairly good agrement with experiments was reported. The influence of hydrostatic pressure on solid nitromethane shows distinct behavior from 8 to 15 GPa. From the weakening of hydrogen bond strength, softening of lattice modes supports a possible structural transition in solid nitromethane between 8 to 12 GPa and this was in excellent agreement with reported experiments. Finally, we have used the GW approximation to obtain an accurate value of the band gap and presented the corresponding quasiparticle band structure. We found that nitromethane
 has an indirect band gap of approximately 7.8 eV which occurs between the high symmetry points $\Gamma$ and X.
The calculated dielectric function, refractive index, and absorption functions along three crystallographic  directions shows a considerable anisotropy
We expect that our results will stimulate further experiments on solid nitromethane but also that they will
 be useful to understand other materials used as explosives.

%The present calculations at high pressures in this study is
%significant to confirms the structural transition and the calculation of quasiparticle band structure gives the first step to predict exact band gaps for secondary
%explosive materials to explore the relation between impact sensitivity and band gap.
%and hope our results could
%stimulate the resistivity as well as optical spectroscopy experiments on solid nitromethane.

\section{Acknowledgments}
S. A. would like to thank DRDO through ACRHEM for financial support and thank CMSD, University of Hyderabad, for providing computational facilities.
S. L acknowledges GENCI-CCRT/CINES (Grant x2013-085106) for access to computational facilities.
$^*$\emph{Author for Correspondence, E-mail: gvaithee@gmail.com}

 \end{document}